\PassOptionsToPackage{unicode}{hyperref}
\PassOptionsToPackage{hyphens}{url}
\PassOptionsToPackage{dvipsnames,svgnames,x11names}{xcolor}
\documentclass[
  article,nojss]{jss}
\usepackage{xcolor}
\usepackage{amsmath,amssymb}
\usepackage{iftex}
\ifPDFTeX
  \usepackage[T1]{fontenc}
  \usepackage[utf8]{inputenc}
  \usepackage{textcomp} % provide euro and other symbols
\else % if luatex or xetex
  \usepackage{unicode-math} % this also loads fontspec
  \defaultfontfeatures{Scale=MatchLowercase}
  \defaultfontfeatures[\rmfamily]{Ligatures=TeX,Scale=1}
\fi
\usepackage{lmodern}
\ifPDFTeX\else
\fi
\IfFileExists{upquote.sty}{\usepackage{upquote}}{}
\IfFileExists{microtype.sty}{% use microtype if available
  \usepackage[]{microtype}
  \UseMicrotypeSet[protrusion]{basicmath} % disable protrusion for tt fonts
}{}
\makeatletter
\@ifundefined{KOMAClassName}{% if non-KOMA class
  \IfFileExists{parskip.sty}{%
    \usepackage{parskip}
  }{% else
    \setlength{\parindent}{0pt}
    \setlength{\parskip}{6pt plus 2pt minus 1pt}}
}{% if KOMA class
  \KOMAoptions{parskip=half}}
\makeatother
\makeatletter
\ifx\paragraph\undefined\else
  \let\oldparagraph\paragraph
  \renewcommand{\paragraph}{
    \@ifstar
      \xxxParagraphStar
      \xxxParagraphNoStar
  }
  \newcommand{\xxxParagraphStar}[1]{\oldparagraph*{#1}\mbox{}}
  \newcommand{\xxxParagraphNoStar}[1]{\oldparagraph{#1}\mbox{}}
\fi
\ifx\subparagraph\undefined\else
  \let\oldsubparagraph\subparagraph
  \renewcommand{\subparagraph}{
    \@ifstar
      \xxxSubParagraphStar
      \xxxSubParagraphNoStar
  }
  \newcommand{\xxxSubParagraphStar}[1]{\oldsubparagraph*{#1}\mbox{}}
  \newcommand{\xxxSubParagraphNoStar}[1]{\oldsubparagraph{#1}\mbox{}}
\fi
\makeatother

\usepackage{longtable,booktabs,array}
\usepackage{calc} % for calculating minipage widths
\usepackage{etoolbox}
\makeatletter
\patchcmd\longtable{\par}{\if@noskipsec\mbox{}\fi\par}{}{}
\makeatother
\IfFileExists{footnotehyper.sty}{\usepackage{footnotehyper}}{\usepackage{footnote}}
\makesavenoteenv{longtable}
\usepackage{graphicx}
\makeatletter
\newsavebox\pandoc@box
\newcommand*\pandocbounded[1]{% scales image to fit in text height/width
  \sbox\pandoc@box{#1}%
  \Gscale@div\@tempa{\textheight}{\dimexpr\ht\pandoc@box+\dp\pandoc@box\relax}%
  \Gscale@div\@tempb{\linewidth}{\wd\pandoc@box}%
  \ifdim\@tempb\p@<\@tempa\p@\let\@tempa\@tempb\fi% select the smaller of both
  \ifdim\@tempa\p@<\p@\scalebox{\@tempa}{\usebox\pandoc@box}%
  \else\usebox{\pandoc@box}%
  \fi%
}
\def\fps@figure{htbp}
\makeatother

\providecommand{\tightlist}{%
  \setlength{\itemsep}{0pt}\setlength{\parskip}{0pt}}

\usepackage{orcidlink,thumbpdf,lmodern}

\usepackage{amsmath}
\usepackage[ruled]{algorithm2e}
\makeatletter
\@ifpackageloaded{caption}{}{\usepackage{caption}}
\AtBeginDocument{%
\ifdefined\contentsname
  \renewcommand*\contentsname{Table of contents}
\else
  \newcommand\contentsname{Table of contents}
\fi
\ifdefined\listfigurename
  \renewcommand*\listfigurename{List of Figures}
\else
  \newcommand\listfigurename{List of Figures}
\fi
\ifdefined\listtablename
  \renewcommand*\listtablename{List of Tables}
\else
  \newcommand\listtablename{List of Tables}
\fi
\ifdefined\figurename
  \renewcommand*\figurename{Figure}
\else
  \newcommand\figurename{Figure}
\fi
\ifdefined\tablename
  \renewcommand*\tablename{Table}
\else
  \newcommand\tablename{Table}
\fi
}
\@ifpackageloaded{float}{}{\usepackage{float}}
\floatstyle{ruled}
\@ifundefined{c@chapter}{\newfloat{codelisting}{h}{lop}}{\newfloat{codelisting}{h}{lop}[chapter]}
\floatname{codelisting}{Listing}

\makeatother
\makeatletter
\@ifpackageloaded{caption}{}{\usepackage{caption}}
\@ifpackageloaded{subcaption}{}{\usepackage{subcaption}}
\makeatother
\makeatletter
\@ifpackageloaded{tcolorbox}{}{\usepackage[skins,breakable]{tcolorbox}}
\makeatother
\makeatletter
\@ifundefined{shadecolor}{\definecolor{shadecolor}{rgb}{.97, .97, .97}}{}
\makeatother
\makeatletter
\ifdefined\Shaded\renewenvironment{Shaded}{\begin{tcolorbox}[enhanced, boxrule=0pt, interior hidden, sharp corners, frame hidden, borderline west={3pt}{0pt}{shadecolor}, breakable]}{\end{tcolorbox}}\fi
\makeatother
\usepackage{bookmark}
\IfFileExists{xurl.sty}{\usepackage{xurl}}{} % add URL line breaks if available
\hypersetup{
  pdftitle={tidysynthesis: a Meta-Package for Synthetic Data Generation},
  pdfauthor={Aaron R. Williams; Jeremy Seeman; Gabriel Morrison},
  colorlinks=true,
  linkcolor={blue},
  filecolor={Maroon},
  citecolor={Blue},
  urlcolor={Blue},
  pdfcreator={LaTeX via pandoc}}

\author{Aaron R. Williams~\orcidlink{0000-0001-5564-1938}\\Urban
Institute \And Jeremy Seeman~\orcidlink{0000-0003-3526-3209}\\Urban
Institute \AND Gabriel Morrison~\orcidlink{0009-0008-1815-5920}\\Urban
Institute}
\Plainauthor{Aaron R. Williams, Jeremy Seeman, Gabriel
Morrison} %% comma-separated

\title{tidysynthesis: a Meta-Package for Synthetic Data Generation}
\Plaintitle{tidysynthesis: a Meta-Package for Synthetic Data
Generation} %% without formatting

\Abstract{Synthetic data generation enables data curators to more easily
share datasets that limits the potential for disclosive inferences about
data subjects in confidential datasets. Generating synthetic data
requires navigating numerous design choices; however, most existing open
source software fails to provide common software infrastructure for
making such design choices efficiently. In this paper, we introduce
\pkg{tidysynthesis}, a meta-package for synthetic data generation that
enables better interoperability between existing modeling frameworks and
statistical data privacy methods. \pkg{tidysynthesis} allows users more
flexibility to specify and iterate on synthetic data algorithms by
providing a common syntax to easily create and modify synthetic data
generation pipelines. We demonstrate the features and extensibility of
\pkg{tidysynthesis}, as well as provide end-to-end examples for
synthetic data generation using data from the American Community
Survey.}

\Keywords{synthetic data, data privacy, R}

\Address{
Aaron R. Williams\\
E-mail: \email{awilliams@urban.org}\\
\\~
Jeremy Seeman\\
E-mail: \email{jseeman@urban.org}\\
\\~
Gabriel Morrison\\
E-mail: \email{gmorrison@urban.org}\\
\\~

}

\begin{document}
\maketitle

\section{Introduction}\label{sec-intro}

\subsection{Synthetic Data and Statistical Data
Privacy}\label{synthetic-data-and-statistical-data-privacy}

Access to high-quality data is essential for building knowledge and
making evidence-based decisions. However, many useful data sets are
generally inaccessible because of disclosure concerns. For example,
\citet{chetty2014} use hundreds of millions of confidential tax returns
to understand inter-generational mobility in the United States; this
high-impact research was only possible through privileged access to
confidential data. Conversely, many existing publicly available datasets
pose substantial disclosure risk concerns, raising questions about how
to improve these processes \citep{henriksen2016re}.

To safely expand access to confidential data, many practitioners are
turning to synthetic data. The term ``synthetic data'' refers to
datasets that imitate the statistical properties of a confidential
datasets while limiting inferences that could be made about confidential
data subjects \citep{raghunathan2021synthetic}. Synthetic data
techniques use generative modeling to draw new synthetic data samples
from models trained on confidential data. When used effectively,
synthetic data offers an alternative data access mechanism for making
more confidential information accessible to researchers or mitigating
disclosure risk for existing data products. For the purposes of this
paper and software, we focus on tabular synthetic data as used in
quantitative social sciences.

Data synthesis has grown in popularity in recent years because it can
leverage randomization to help limit membership inferences, i.e.,
whether an adversary can infer the presence or absence of a particular
record in the confidential data. With increasing computing power,
third-party data availability, and auxiliary information, traditional
techniques like suppression, swapping, and bottom- and top-coding suffer
from exposure to membership inference with certainty
\citep{machanavajjhala2007diversity, cohen2022attacks}. Moreover, these
same traditional techniques make it harder for data users to produce
accurate inferences, as the mechanisms use to modify the confidential
data cannot be easily corrected \citep{fienberg2004data}. Organizations
like the U.S. Census Bureau \citep{benedetto2013, kinney2011} and the
Internal Revenue Service \citep{bowen2022} are adopting synthetic data
products to address these challenges.

Data curators considering synthetic data products must navigate the
unavoidable trade-off between privacy (i.e., the degree to which
disclosure risks are mitigated) and utility (i.e., the degree to which
the resulting data products are useful for statistical analyses)
\citep{dinur2003revealing}. We will refer to this as the
``privacy-utility trade-off.'' Synthetic data methods and their
implementations span the full spectrum of this trade-off, because as an
umbrella term, ``synthetic data'' can refer to \emph{any} generative
model output that imitates the structural or schematic properties of a
confidential dataset. Some synthetic data techniques may overfit to the
confidential data, posing disclosure risk concerns. Alternatively, other
synthetic data techniques may underfit to the confidential data,
diminishing the usefulness of the synthetic data.

\subsection{Contributions}\label{contributions}

Effective synthetic data implementations require exploring diverse
methodologies and frameworks for synthetic data generation. However,
there's little common infrastructure in R's open-source software
ecosystem to support this exploration. We developed
\texttt{tidysynthesis}, a meta-package for synthetic data generation, to
address precisely this problem. In this paper, we outline the following
contributions:

\begin{enumerate}
\def\labelenumi{\arabic{enumi}.}
\tightlist
\item
  In Section~\ref{sec-method}, we provide background information about
  synthetic data methodology and discuss the need for standardizing
  common synthetic data design decisions in the current software
  environment.
\item
  In Section~\ref{sec-package}, we introduce \texttt{tidysynthesis}, our
  meta-package. We discuss its design principles and outline a subset of
  its features.
\item
  In Section~\ref{sec-example}, we use \texttt{tidysynthesis} to
  generate a synthetic sample from the American Community Survey
  \citep{census_acs_pums_2019}.
\end{enumerate}

\section{Methodological Background and Motivation}\label{sec-method}

\subsection{Synthetic Data Generation and Design
Decisions}\label{synthetic-data-generation-and-design-decisions}

Responsible quantitative research about people or organizations aims to
make inferences about populations and subpopulations without enabling
disclosive inferences about the subjects contributing sensitive data.
Enabling population-level inferences while limiting unit-level
inferences is the core problem of ``statistical data privacy'' (SDP)
\citep{slavkovic2023statistical}, which encompasses many data privacy
interventions and measurement techniques from theoretical and applied
computer science, statistics, and other related fields. Frameworks like
``statistical disclosure control or limitation''
\citep{hundepool2012statistical} and ``differential privacy''
\citep{dwork2014algorithmic} are all associated with SDP, and each has
different methods for generating and evaluating synthetic data. For
example, differentially private synthetic data requires the algorithmic
form of any randomization injected into the data generating process to
have specific mathematical properties \citep{dwork2014algorithmic}.
Practitioners must be able to make comparisons based on different SDP
frameworks and implementations to understand how different synthetic
data methods have different privacy and utility implications.

For the purposes of this paper and this software, we focus on fully
conditional specifications (FCS) for generating synthetic data
\citep{reiter2005using}. FCS uses a sequential approach to model the
joint distribution of a confidential data set as a sequence of
univariate conditional distributions. This approach enables sequential
modeling of confidential data variables with a high level of
user-specified variability in model properties across conditional
distributions while remaining mathematically equivalent with the full
joint generative models. FCS also serves a pragmatic purpose by breaking
down memory-intensive model fitting processes into sequential processing
steps, ultimately decreases memory demands while increasing total
computation time.

To illustrate FCS, suppose we want to generate a synthetic dataset with
\(p\) tabular variables. Let \(X_k\) refer to the \(k\)th random
variable that describes the \(k\)th tabular variable, and let
\(X_{1:k}\) refer to the joint random variable of the first \(k\)
tabular variables. Recall that by the classical definition of
conditional probability, for any \(s\) in \(\{ 1, 2, \dots, p - 1 \}\),

\[
P(X_{1:p}) = P(X_{1:s}) P(X_{s+1} \mid X_{1:s}) \ P(X_{s+2} \mid X_{1:s+1}) \dots P(X_{p} \mid X_{1:p-1})
\]

FCS operates under this decomposition principle by using the following
general-purpose algorithm for generating synthetic data. To generate
\(n\) synthetic data records with FCS, we use the following steps. Let
\(\hat{\theta}_{1:k}\) refer to modeling parameters for the first \(k\)
tabular variables (with conditional notation defined similarly).

\begin{algorithm}
\caption{Fully Conditional Synthetic Data Algorithm}
\SetKwInOut{Input}{Input}
\SetKwInOut{Output}{Output}
\Input{Confidential data: $X$}
\Input{Starting Model Parameters: $\hat{\theta}_{1:s}$}
\Input{Starting Probability Model: $P_{\hat{\theta}_{1:s}}$}
\Input{Sequential Model Parameters: $\hat{\theta}_{k|1:(k-1)}, k \in \{ s+1, s+2, \dots, p \}$}
\Input{Sequential Probability Models: $P_{\hat{\theta}_{k|1:(k-1)}}, k \in \{ s+1, s+2, \dots, p \}$}
\Output{Synthetic data: $Y_{1,1:p}, \dots, Y_{n,1:p}$}
Compute $\hat{\theta}_{1:s}(X)$\;
Sample $Y_{1,1:s}, \dots, Y_{n,1:s} \sim P_{\hat{\theta}_{1:s}}(\cdot)$ \;
\For{$k \in \{ s+1, s+2, \dots, p\} $}{
  Compute $\hat{\theta}_{k|1:(k-1)}(X)$ \;
  Sample $Y_{1,k}, \dots, Y_{n,k} \sim P_{\hat{\theta}_k}(\cdot \mid Y_{1,1:k-1}, \dots, Y_{n,1:k-1})$ \;
}
\end{algorithm}

Implementing the generic algorithm above requires making the following
decisions:

\begin{itemize}
\tightlist
\item
  \textbf{Visit sequence}: what tabular variables correspond to which
  column-wise indices in all the steps above? In other words, what is
  \(s\), which variables get synthesized in the initial batch of \(s\)
  variables, and in what order do the remaining variables get
  synthesized?
\item
  \textbf{Models}: how do we determine \(\hat{\theta}_k\) and
  \(\hat{\theta}_{k | 1:(k-1)}\) for each
  \(k \in \{s+1, s+2, \dots, n \}\)? In other words, how do we fit
  models that characterize the distributions for each component of the
  FCS?
\item
  \textbf{Samplers}: how do we determine \(P_{\hat{\theta}_k}\) and
  \(P_{\hat{\theta}_{k | 1:(k-1)}}\) for each
  \(k \in \{s+1, s+2, \dots, n \}\)? In other words, how do we sample
  new synthetic values from each modeled distribution?
\end{itemize}

Generative models that navigate the privacy-utility trade-off require
making numerous design choices outlined in Algorithm 1 above, and the
number of design choices can quickly become untenable in
higher-dimensional settings. Responsible synthetic data usage
necessarily requires some degree of empirical experimentation that
compares synthetic data implementations against one another, regardless
of which metrics are used to evaluate the synthetic data.

\subsection{Related Work: Common Infrastructure for Synthetic Data
Generation}\label{related-work-common-infrastructure-for-synthetic-data-generation}

Users would ideally be able to modify one of these many design decisions
and easily see how their synthetic data changes. However, the existing R
package ecosystem fails to support this necessary experimentation for
two main reasons. First, most synthetic data generation packages are
tied to specific generative model methodologies that are inflexible by
design. These packages are typically issued by the authors of papers
proposing the methodology, meaning each is a priori opinionated about
the ``best'' way to synthesize data. Second, many popular synthetic data
packages are not easily interoperable with common data science workflows
for generative model fitting and sampling.

To demonstrate these issues, we look at all packages currently hosted on
CRAN that enable tabular synthetic data generation. We filter CRAN
results based on those which explicitly mention ``synthetic data'' in
their description, and we exclude packages which use
domain-knowledge-based simulation (for example, climate data simulated
from mechanical models), synthetic control methods in causal inference
(for example, synthesizing treatment comparison groups), or non-tabular
data (for example, synthesizing graph data). The following packages are
displayed in Table~\ref{tbl-cran}. This list is by no means exhaustive,
as there are many synthetic data tools which may not be hosted on CRAN
or do not explicitly mention synthetic data.

\begin{longtable}[]{@{}
  >{\raggedright\arraybackslash}p{(\linewidth - 4\tabcolsep) * \real{0.1300}}
  >{\raggedright\arraybackslash}p{(\linewidth - 4\tabcolsep) * \real{0.6400}}
  >{\raggedright\arraybackslash}p{(\linewidth - 4\tabcolsep) * \real{0.2300}}@{}}

\toprule\noalign{}
\begin{minipage}[b]{\linewidth}\raggedright
Package
\end{minipage} & \begin{minipage}[b]{\linewidth}\raggedright
URL
\end{minipage} & \begin{minipage}[b]{\linewidth}\raggedright
General.Method.Type
\end{minipage} \\
\midrule\noalign{}
\endhead
\bottomrule\noalign{}
\endlastfoot
arf & https://cran.r-project.org/web/packages/arf/index.html & Black-Box
Optimization \\
conjurer & https://cran.r-project.org/web/packages/conjurer/index.html &
Parametric Models \\
FakeDataR & https://cran.r-project.org/web/packages/FakeDataR/index.html
& Black-Box Optimization \\
humanleague &
https://cran.r-project.org/web/packages/humanleague/index.html &
Resampling \\
LikertMakeR &
https://cran.r-project.org/web/packages/LikertMakeR/index.html &
Parametric Models \\
modgo & https://cran.r-project.org/web/packages/modgo/index.html &
Parametric Models \\
PopulateR & https://cran.r-project.org/web/packages/PopulateR/index.html
& Resampling \\
RealSurvSim &
https://cran.r-project.org/web/packages/RealSurvSim/index.html &
Nonparametric Models \\
RGAN & https://cran.r-project.org/web/packages/RGAN/index.html &
Black-Box Optimization \\
simPop & https://cran.r-project.org/web/packages/simPop/index.html &
Parametric Models \\
synMicrodata &
https://cran.r-project.org/web/packages/synMicrodata/index.html &
Nonparametric Models \\
synthesis & https://cran.r-project.org/web/packages/synthesis/index.html
& Parametric Models \\
synthesizer &
https://cran.r-project.org/web/packages/synthesizer/index.html &
Resampling \\
synthpop & https://cran.r-project.org/web/packages/synthpop/index.html &
Black-Box Optimization \\

\caption{\label{tbl-cran}A list of CRAN synthetic data packages.}

\tabularnewline
\end{longtable}

\textsubscript{Source:
\href{https://UI-Research.github.io/tidysynthesis-paper/tidysynthesis.qmd.html}{Article
Notebook}}

Each package in Table~\ref{tbl-cran} prescribes either one specific
model implementation or a limited family of model implementations. These
implementations span a wide variety of methodologies with different
parameterization assumptions, estimation procedures, and probability
interpretations. With few exceptions, most of these packages require the
prescribed modeling approach to be applied to all variables
simultaneously and in-memory. This limits the ability for users to
customize synthesis models for specific variables or iteratively build
large synthetic datasets when memory is limited.

Moreover, there are countless implementations for conditional models
that are not interoperable with these libraries. FCS enables using
discriminative models for generative purposes, where predictions
generated from discriminative models form the basis for new generative
samples. By specifying how one samples from discriminative models, they
can behave like generative models by iteratively adding new variables to
synthetic records. Many existing CRAN packages support simplistic
integrations with base R functionality like \texttt{stats::glm}, but
only a small handful of modern techniques are supported. Notably, these
packages exclude broader modeling ecosystems like those in
\texttt{tidymodels}.

\section{Our Proposed Solution: tidysynthesis}\label{sec-package}

For these reasons, we developed \texttt{tidysynthesis} as a meta-package
to support synthetic data generative modeling experimentation with
interoperability with \texttt{tidyverse} packages (and, consequentially,
any modeling pipeline that can be represented using \texttt{tidymodels}
syntax). \texttt{tidysynthesis} reflects a design philosophy informed by
the tidy tools manifesto\footnote{https://cran.r-project.org/web/packages/tidyverse/vignettes/manifesto.html},
tidymodels design document\footnote{https://tidymodels.github.io/model-implementation-principles/},
and popular ``tidy'' open-source R packages. We highlight a few major
design decisions that the rest of this paper will explain and
demonstrate.

\begin{enumerate}
\def\labelenumi{\arabic{enumi}.}
\tightlist
\item
  \texttt{tidysynthesis} is built on top of the \texttt{tidymodels}
  framework \citep{kuhnaut2023}. \texttt{tidymodels} is a ``collection
  of packages for modeling and statistical analysis that share the
  underlying design philosophy, grammar, and data structures of the
  tidyverse.'' \texttt{tidymodels} provides a common interface to most
  predictive models in R and complete tools for feature and target
  engineering. Sequential synthesis is only as good as the models used
  to predict values for each variable, and \texttt{tidymodels} offers a
  complete predictive modeling framework.
\item
  \texttt{tidysynthesis} is open architecture software. The package is
  modular and transparent such that users can create their own functions
  and plug them into existing functions to extend or change the
  functionality of \texttt{tidysynthesis}.
\item
  \texttt{tidysynthesis} embraces small, clear functions over large
  functions with many arguments. These functions are like building
  blocks that users can put together into many different synthesis
  specifications. This design decision was directly motivated by changes
  made between \texttt{caret} \citep{kuhnaut2023a} and its successor
  package \texttt{tidymodels}.
\item
  \texttt{tidysynthesis} aims to separate the specification and
  computation of a synthesis. This makes it easier to reuse objects.
  More importantly, it makes it easier to catch mistakes before
  computation, which can be computationally expensive. For example, the
  goal is to catch the mismatch between a variable type and a model type
  during specification instead of in middle of a computationally
  expensive synthesis.
\end{enumerate}

Additionally, \texttt{tidysynthesis} includes functionality not
available in other popular synthesis packages like \texttt{synthpop}.
The package contains robust tools for feature and target engineering,
hyperparameter tuning, interfaces for a broader range of generative
models, a novel framework for injecting additional noise into sampled
values, and the ability to add and enforce mid-synthesis constraints. We
believe these design decisions make \texttt{tidysynthesis} the most
flexible and powerful tool for sequential synthesis.

\texttt{tidysynthesis} offers a simple interface to design, compare, and
iterate on synthetic data model specifications. Instead of writing
separate code to fit each model and use each model to generate samples,
\texttt{tidysynthesis} provides a meta-syntax to specify sequences of
these actions with the flexibility to change the behavior for any subset
of variables. Most importantly, each required and optional component of
a synthesis is fully extensible, so innovations in data synthesis can be
added to the package as the field matures.

\subsection{Obtaining the Software}\label{obtaining-the-software}

\texttt{tidysynthesis} is available on CRAN
\href{https://cran.r-project.org/web/packages/tidysynthesis/index.html}{here},
and the source code is available on GitHub
\href{https://github.com/UrbanInstitute/tidysynthesis}{here}.

\begin{verbatim}
# installing `tidysynthesis` from CRAN
install.packages("tidysynthesis")

# installing `tidysynthesis` from GitHub
pak::pak("UrbanInstitute/tidysynthesis")
\end{verbatim}

\textsubscript{Source:
\href{https://UI-Research.github.io/tidysynthesis-paper/tidysynthesis.qmd.html}{Article
Notebook}}

\subsection{Basic Functionality}\label{basic-functionality}

Syntheses with \texttt{tidysynthesis} are organized with a series of
functions that create objects using R's S3 object system. Each object
has an associated constructor of the same name; for example, the
function \texttt{roadmap(...)} produces a object of class
\texttt{roadmap}.

\begin{figure}

\centering{

\includegraphics[width=3in,height=2.55in]{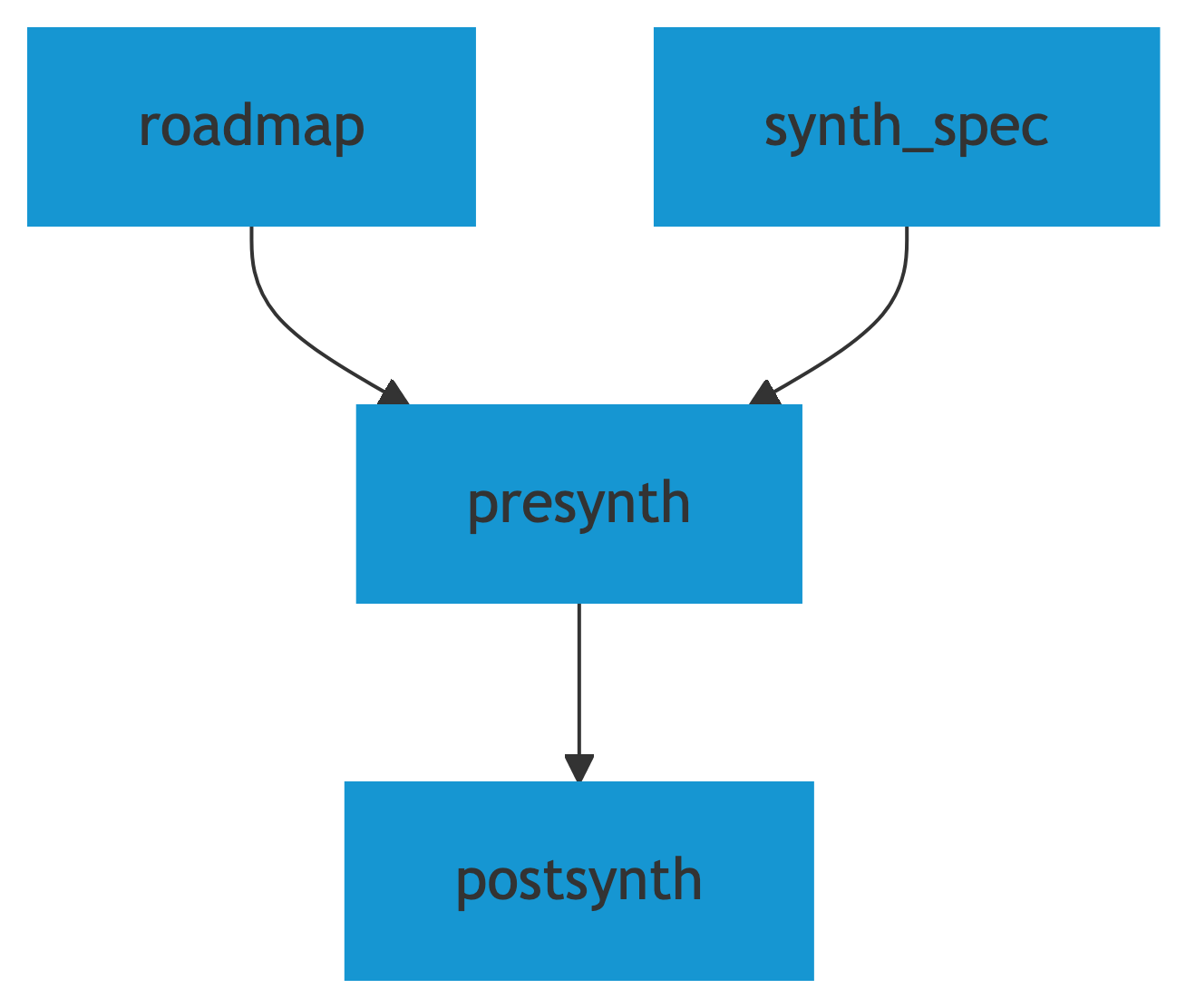}

}

\caption{\label{fig-components-overall}A diagram of the mains
\texttt{tidysynthesis} components.}

\end{figure}

Figure~\ref{fig-components-overall} shows the two primary objects in
\texttt{tidysynthesis}, \texttt{roadmap} and the \texttt{synth\_spec}.
These are the required inputs to \texttt{presynth}, which contains
everything necessary to run a synthesis without actually executing the
process. By calling the \texttt{synthesize()} function on a
\texttt{presynth} object, \texttt{tidysynthesis} returns a
\texttt{postsynth} object containing the resulting synthetic data and
runtime properties. Note that \texttt{presynth} objects perform
extensive error checking when they are created. For example,
\texttt{presynth()} ensures that every variable in the visit sequence
has an appropriate model and sampler. This aims to mitigate downstream
errors in syntheses, which can be time-consuming to correct.

\textsubscript{Source:
\href{https://UI-Research.github.io/tidysynthesis-paper/tidysynthesis.qmd.html}{Article
Notebook}}

\begin{figure}

\centering{

\includegraphics[width=6.5in,height=0.88in]{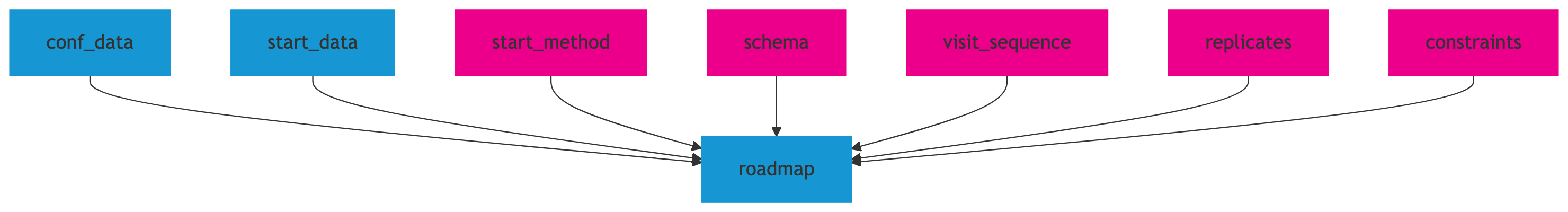}

}

\caption{\label{fig-components-roadmap}A diagram of \texttt{roadmap}
components. Objects in blue are required while objects in magenta are
optional.}

\end{figure}

The \texttt{roadmap} object describes input data sources and macroscopic
properties about the synthesis order of operations. \texttt{roadmap}
instances require two arguments, the confidential data
\texttt{conf\_data} and starting data \texttt{start\_data} (used for the
initial generative model). All remaining arguments are optional and
described in more detail below and are included in
Figure~\ref{fig-components-roadmap}.

\textsubscript{Source:
\href{https://UI-Research.github.io/tidysynthesis-paper/tidysynthesis.qmd.html}{Article
Notebook}}

\begin{figure}

\centering{

\includegraphics[width=6.5in,height=1.21in]{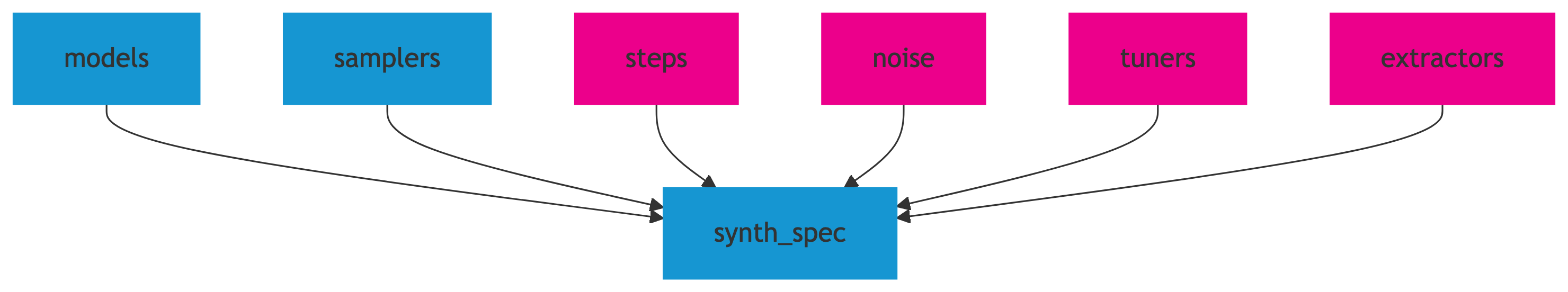}

}

\caption{\label{fig-components-synth_spec}A diagram of
\texttt{synth\_spec} components. Objects in blue are required while
objects in magenta are optional.}

\end{figure}

\texttt{synth\_spec} specifies the generative modeling and sampling
procedures needed for different conditional models in the sequential
synthesis. \texttt{synth\_spec} instances require models (specified
using the \texttt{parsnip::model\_spec} convention) and samplers
(functions with a shared signature defined in the package). We
distinguish between modeling and sampling steps because in either stage,
we may deviate from traditional generative modeling techniques (for
example, using additional privacy-preserving randomization at the
modeling or sampling stage). Models and samplers can be specified based
on output type (numeric or categorical) as well as through custom
mappings between variable names and model or sampler objects. All
remaining arguments are optional and described in more detail below and
are included in Figure~\ref{fig-components-synth_spec}.

\textsubscript{Source:
\href{https://UI-Research.github.io/tidysynthesis-paper/tidysynthesis.qmd.html}{Article
Notebook}}

\subsection{Additional Functionality}\label{additional-functionality}

\texttt{tidysynthesis} includes additional functionality for generating
synthetic data. The following arguments can be optionally supplied to
\texttt{roadmap} or \texttt{synth\_spec}.

\texttt{start\_method}: by default, the \texttt{start\_data} object is
used unmodified as the initial variables in the resulting synthetic
data. If \texttt{start\_data} contains exact confidential records, this
would pose disclosure risks. The \texttt{start\_method} object lets
users supply a (typically randomized) function that modifies the
starting data. This allows users to specify generative modeling
procedures for the starting data, from simple approaches like resampling
to more complex approaches like self-contained generative modeling
calls.

\texttt{schema}: by default, information about variables (i.e., tabular
data columns) are inferred from \texttt{conf\_data}. Users may
optionally specify \texttt{schema} objects which contain information
about variable types, missingness specifications, factor level
definitions, and modes for enforcing these properties both on
confidential data and resulting synthetic data.

\texttt{visit\_sequence}: the visit sequence determines the order in
which variables are included in the sequential synthesis model. By
default, the \texttt{visit\_sequence} is pre-specified using the order
in which variables appear in the confidential data. However, this is
seldom an optimal approach, since variables placed later in the visit
sequence are able to use more previous variables as predictors. In
addition to letting users specify custom \texttt{visit\_sequence}
instances, we provide helper functions to modify the visit sequence
based on empirical metrics from the confidential data (examples below).

\texttt{replicates}: by default, a \texttt{postsynth} contains only one
synthetic data replicate. Alternatively, users can specify a
\texttt{replicates} object which allows users to repeat part or all of
the synthesis process multiple times. This can help users perform
empirical uncertainty quantification, either end-to-end or isolating
different points in the synthesis process.

\texttt{constraints}: by default, synthesized numeric variables can take
any real value, and synthesized categorical values can take any value in
the variable's support. However, it's common for synthesized variables
to obey constraints on allowed values, such as those that globally apply
(for example, certain numeric variables are strictly non-negative) or
conditionally apply (for example, variables that must take specific
values contingent on other synthesized values). The \texttt{constraints}
object lets users specify these constraints for both numeric and
categorical variables. Following \citet{drechsler2011}, constraints may
be enforced either by rejection sampling (i.e., drawing new synthetic
values and only accepting those which conform to specified constraints),
and/or by explicit post-processing (i.e., projecting numeric variables
to the upper or lower bounds of specified constraints, should rejection
sampling fail to produce acceptable records after a fixed number of
attempts). More details are available in the examples below.

\texttt{steps}: by default, each generative model will use each
predictor variable available to it ``as-is''; the available predictor
variables will be defined in an automatically populated \texttt{recipe}.
Users may want to specify feature engineering procedures, variable
inclusion criteria, or other preprocessing steps prior to modeling. To
do this, users can specify \texttt{steps} which perform these
pre-processing tasks within the \texttt{library(recipes)} framework.

\texttt{noise}: by default, each sample from the generative model is
returned as-is when appended to the working synthetic data. Users may
choose to inject additional randomization to further modify these
samples (for example, if the ``samples'' are actually conditional
predicted values). The \texttt{noise} function lets users specify
data-dependent and data-independent noise mechanisms for increasing the
variability of synthesized values.

\texttt{tuners}: by default, each model uses the default hyperparameter
settings supplied to it from the original \texttt{parsnip}
specification. Some modeling procedures may have hyperparameters that
users may wish to set in a data-dependent manner. Users can specify
tuning arguments here for use in \texttt{library(tune)}, allowing
certain hyperparameters of the underlying \texttt{parsnip} models to be
empirically optimized.

\texttt{extractors}: by default, no generative model fitting outputs are
accessible once a synthesis has been completed. However, users may
provide optional \texttt{extractor} functions to return specific
information from models once they've been fit.

\texttt{tidysynthesis} supports user-friendly and iterative development
of synthetic data. The package is designed to catch errors before time-
and compute-intensive sequential synthesis operations begin. Even so,
calls to \texttt{synthesize()} may still fail mid-synthesis either due
to users mis-specifying some component of the synthesis process or as a
result of failures in the compute environment unrelated to
\texttt{tidysynthesis}. This can pose challenges for package users,
especially when working with large data or in compute-constrained
environments. To mitigate this, \texttt{synthesize()} optionally allow
users to save all synthetic data generated up to the variable on which
the synthesis fails. In addition, the package includes two functions
\texttt{postsynth\_to\_roadmap()} and
\texttt{postsynth\_to\_synth\_spec()} that take a partially completed
\texttt{postsynth} object, update it based on which variables were
synthesized successfully, and return a new \texttt{roadmap} or
\texttt{synth\_spec} to be used in subsequent syntheses. Using these
functions may make reproducing exact synthetic data-sets difficult and
are therefore ill-suited to creation a ``final'' version of synthetic
data. Nonetheless, they can reduce the time it takes to test
specifications and resolve bugs, import parts of the synthetic data
generation process in practice.

\section{Examples}\label{sec-example}

The following section uses \texttt{tidysynthesis} to generate synthetic
datasets based on a public use microdata sample from the American
Community Survey. The extract is accessed through IPUMS
\citep{ruggles2025IPUMS} and generated using \texttt{library(ipumsr)};
the data is available in \texttt{tidysynthesis}.

\subsection{Data}\label{data}

The American Community Survey (ACS) is an essential source of
demographic and economic information in the United States. We select a
disjoint sample from the 2019 1-year ACS responses, one with 1500
samples (\texttt{acs\_conf}).

The table below lists the available variables:

\begin{longtable}[]{@{}
  >{\raggedright\arraybackslash}p{(\linewidth - 4\tabcolsep) * \real{0.1687}}
  >{\raggedright\arraybackslash}p{(\linewidth - 4\tabcolsep) * \real{0.1566}}
  >{\raggedright\arraybackslash}p{(\linewidth - 4\tabcolsep) * \real{0.6747}}@{}}
\toprule\noalign{}
\begin{minipage}[b]{\linewidth}\raggedright
VariableName
\end{minipage} & \begin{minipage}[b]{\linewidth}\raggedright
VariableType
\end{minipage} & \begin{minipage}[b]{\linewidth}\raggedright
Description
\end{minipage} \\
\midrule\noalign{}
\endhead
\bottomrule\noalign{}
\endlastfoot
\texttt{county} & categorical & County of residence \\
\texttt{gq} & categorical & Kind of residence (single family, group
quarters, etc.) \\
\texttt{sex} & categorical & Head-of-household sex \\
\texttt{marst} & categorical & Head-of-household marital status \\
\texttt{empstat} & categorical & Head-of-household employment status \\
\texttt{age} & numeric & Head-of-household age \\
\texttt{famsize} & numeric & Number of residents in household \\
\texttt{transittime} & numeric & Minutes of transit time for
head-of-household to work \\
\texttt{inctot} & numeric & Total annual household income \\
\end{longtable}

\textsubscript{Source:
\href{https://UI-Research.github.io/tidysynthesis-paper/tidysynthesis.qmd.html}{Article
Notebook}}

\subsection{Basic Examples}\label{basic-examples}

First we load the necessary packages and data.

\begin{verbatim}
library(tidyverse)
\end{verbatim}

\begin{verbatim}
library(tidysynthesis)

# load ACS data and select confidential data variables
acs_conf_data <- tidysynthesis::acs_conf |>
  select(county, gq, sex, marst, empstat, age, famsize, transit_time, inctot)

glimpse(acs_conf_data)
\end{verbatim}

\begin{verbatim}
Rows: 1,500
Columns: 9
$ county       <fct> Other, Other, Other, Other, Douglas, Lancaster, Other, Sa~
$ gq           <fct> Household, Household, Household, Household, Household, Ho~
$ sex          <fct> Female, Male, Male, Female, Male, Female, Male, Male, Mal~
$ marst        <fct> Single, Married, Single, Single, Married, Divorced, Marri~
$ empstat      <fct> NA, Employed, NA, NA, Employed, Employed, NA, NA, NA, Emp~
$ age          <dbl> 0, 41, 10, 12, 46, 36, 49, 5, 22, 31, 5, 55, 74, 50, 37, ~
$ famsize      <dbl> 5, 4, 3, 6, 5, 3, 5, 5, 4, 1, 4, 2, 2, 2, 4, 1, 1, 4, 5, ~
$ transit_time <dbl> 0, 30, 0, 0, 15, 15, 0, 0, 0, 5, 0, 7, 0, 15, 10, 0, 0, 0~
$ inctot       <dbl> NA, 68000, NA, NA, 91000, 26200, 6000, NA, 0, 37000, NA, ~
\end{verbatim}

\textsubscript{Source:
\href{https://UI-Research.github.io/tidysynthesis-paper/tidysynthesis.qmd.html}{Article
Notebook}}

For this example, we will use the first four categorical variables as
our starting data. This illustration will use the same data as above,
but \texttt{tidysynthesis} supports any start data whose names and data
types match a subset of confidential data variables.

\begin{verbatim}
# select starting variables 
acs_start_data <- acs_conf_data |>
  select(county, gq, sex, marst)

glimpse(acs_start_data)
\end{verbatim}

\begin{verbatim}
Rows: 1,500
Columns: 4
$ county <fct> Other, Other, Other, Other, Douglas, Lancaster, Other, Sarpy, O~
$ gq     <fct> Household, Household, Household, Household, Household, Househol~
$ sex    <fct> Female, Male, Male, Female, Male, Female, Male, Male, Male, Fem~
$ marst  <fct> Single, Married, Single, Single, Married, Divorced, Married, Si~
\end{verbatim}

\textsubscript{Source:
\href{https://UI-Research.github.io/tidysynthesis-paper/tidysynthesis.qmd.html}{Article
Notebook}}

We supply these two datasets as required arguments to create a
\texttt{roadmap} instance. For the rest of this paper, we will
manipulate this \texttt{roadmap} using other \texttt{tidysynthesis}
functionality.

\begin{verbatim}
# create a minimal `roadmap` instance
roadmap_acs <- roadmap(
  conf_data = acs_conf_data,  # confidential data (matches synthetic data)
  start_data = acs_start_data  # start data (for first generative model)
)

roadmap_acs
\end{verbatim}

\begin{verbatim}
Roadmap: 
conf_data: 1500 observations, 9 variables 
start_data: 1500 observations, 4 variables
\end{verbatim}

\textsubscript{Source:
\href{https://UI-Research.github.io/tidysynthesis-paper/tidysynthesis.qmd.html}{Article
Notebook}}

Next, we will demonstrate ways in which we can update our
\texttt{roadmap} object. The current roadmap uses \texttt{start\_data}
as-is without modification (i.e., producing partially synthetic data).
We can update the \texttt{start\_method} to resample records with
replacement from the original starting data:

\begin{verbatim}
# update start_method 
roadmap_acs <- roadmap_acs |>
  update_start_method(  # update how start_data enters the synthesis process
    start_func = start_resample,  # resample records with replacement
    n = 500  # modify our start_data to sample only 500 records
  )

roadmap_acs[["start_method"]]
\end{verbatim}

\begin{verbatim}
Start Method: User-Specified 
Keyword Arguments: 
n: 500
\end{verbatim}

\textsubscript{Source:
\href{https://UI-Research.github.io/tidysynthesis-paper/tidysynthesis.qmd.html}{Article
Notebook}}

Next, we look at the \texttt{visit\_sequence}. By default, variables are
synthesized in the order in which they appear in the confidential data:

\begin{verbatim}
roadmap_acs[["visit_sequence"]]
\end{verbatim}

\begin{verbatim}
Visit Sequence
Method:Variable
default:empstat default:age default:famsize default:transit_time default:inctot 
\end{verbatim}

\textsubscript{Source:
\href{https://UI-Research.github.io/tidysynthesis-paper/tidysynthesis.qmd.html}{Article
Notebook}}

We can build a \texttt{visit\_sequence} using both manual specifications
and data-driven specifications. Below, we'll specify the sequence by
ordering numeric variables by decreasing proportion of non-zero values.

\begin{verbatim}
# update visit_sequence
roadmap_acs <- roadmap_acs |> 
  add_sequence_numeric( # update the visit_sequence for numeric variables
    dplyr::where(is.numeric), # for each numeric variable in the sequence...
    method = "proportion", # order by decreasing proportion of non-zero values...
    na.rm = TRUE # ...while ignoring NA values
  )

roadmap_acs[["visit_sequence"]]
\end{verbatim}

\begin{verbatim}
Visit Sequence
Method:Variable
proportion:famsize proportion:age proportion:inctot proportion:transit_time default:empstat 
\end{verbatim}

\textsubscript{Source:
\href{https://UI-Research.github.io/tidysynthesis-paper/tidysynthesis.qmd.html}{Article
Notebook}}

Next, we create modeling algorithms from \texttt{library(tidymodels)}
for synthesizing each variable. We'll use classical CART-based
synthesis, so decision trees for categorical variables and regression
trees for numeric variables \citep{breiman_classification_2017}. In both
cases, we'll use the \texttt{rpart} implementation \citep{therneau2022}
accessed through \texttt{library(parsnip)}.

\begin{verbatim}
# specify an example decision tree model for categorical variables
rpart_mod_cat <- parsnip::decision_tree(
  cost_complexity = 0.0001, 
  min_n = 20
) |> 
  parsnip::set_mode(mode = "classification") |>
  parsnip::set_engine(engine = "rpart")

# specify an example decision tree model for regression variables
rpart_mod_num <- parsnip::decision_tree(  
  cost_complexity = 0.0001, 
  min_n = 20
) |>
  parsnip::set_mode(mode = "regression") |>
  parsnip::set_engine(engine = "rpart")
\end{verbatim}

\textsubscript{Source:
\href{https://UI-Research.github.io/tidysynthesis-paper/tidysynthesis.qmd.html}{Article
Notebook}}

Additionally, we create a custom regression tree for the \texttt{age}
variable with slightly different hyperparameters.

\begin{verbatim}
# specify a special example decision tree model for the `age` variable
rpart_mod_num_custom <- parsnip::decision_tree(  
  cost_complexity = 0.01, 
  min_n = 30
) |>
  parsnip::set_mode(mode = "regression") |>
  parsnip::set_engine(engine = "rpart")
\end{verbatim}

\textsubscript{Source:
\href{https://UI-Research.github.io/tidysynthesis-paper/tidysynthesis.qmd.html}{Article
Notebook}}

Next, we need to assign an algorithm to each variable. We'll set
\texttt{rpart\_mod\_cat} as the default model for classification and
\texttt{rpart\_mod\_num} as the default model for regression.
Additionally, we specify the custom model for the \texttt{age} variable.
Finally, we will use the \texttt{tidysynthesis} function
\texttt{sample\_rpart()} to draw new predicted values from each model.

\begin{verbatim}
# create a synthesis specification
synth_spec_acs <- synth_spec(
  # default models by regression and classification
  default_regression_model = rpart_mod_num, 
  default_classification_model = rpart_mod_cat, 
  # associate `age` with our custom model
  custom_models = list(
    list(vars = c("age"), model = rpart_mod_num_custom)
  ),
  # use tidysynthesis-provided sampler function for `rpart`
  default_regression_sampler = sample_rpart,
  default_classification_sampler = sample_rpart
)

synth_spec_acs
\end{verbatim}

\begin{verbatim}
Synthesis specification with user-specified components: 
* default_regression_model
* default_classification_model
* custom_models
* default_regression_sampler
* default_classification_sampler
\end{verbatim}

\textsubscript{Source:
\href{https://UI-Research.github.io/tidysynthesis-paper/tidysynthesis.qmd.html}{Article
Notebook}}

We combine the \texttt{roadmap} and \texttt{synth\_spec} above into a
single \texttt{presynth} object, and then we can \texttt{synthesize()}
the data:

\begin{verbatim}
# construct a `presynth` with a `roadmap` and `synth_spec`
presynth1 <- presynth(
  roadmap = roadmap_acs, 
  synth_spec = synth_spec_acs
)
\end{verbatim}

\begin{verbatim}
# perform a basic synthesis
result1 <- synthesize(presynth1, progress = TRUE)
\end{verbatim}

\textsubscript{Source:
\href{https://UI-Research.github.io/tidysynthesis-paper/tidysynthesis.qmd.html}{Article
Notebook}}

\begin{verbatim}
glimpse(result1[["synthetic_data"]])
\end{verbatim}

\begin{verbatim}
Rows: 500
Columns: 10
$ county       <fct> Other, Sarpy, Other, Lancaster, Lancaster, Other, Other, ~
$ gq           <fct> Household, Household, Household, Household, Household, Ho~
$ sex          <fct> Male, Male, Female, Female, Female, Female, Female, Femal~
$ marst        <fct> Married, Single, Married, Married, Married, Married, Divo~
$ famsize      <dbl> 2, 4, 3, 5, 2, 3, 2, 2, 2, 1, 2, 1, 3, 3, 3, 4, 1, 3, 1, ~
$ age          <dbl> 59, 40, 35, 38, 68, 51, 75, 24, 49, 19, 75, 45, 40, 4, 41~
$ inctot_NA    <fct> nonmissing value, nonmissing value, nonmissing value, non~
$ inctot       <dbl> 45000, 12000, 30000, 75000, 7000, 26180, 56000, 45000, 65~
$ transit_time <dbl> 5, 0, 15, 10, 0, 28, 20, 3, 10, 0, 0, 0, 0, 0, 20, 20, 20~
$ empstat      <fct> Employed, NA, Employed, Employed, NA, Employed, Employed,~
\end{verbatim}

\textsubscript{Source:
\href{https://UI-Research.github.io/tidysynthesis-paper/tidysynthesis.qmd.html}{Article
Notebook}}

\subsection{Additional Features}\label{additional-features}

We briefly demonstrate a few additional features offered by
\texttt{tidysynthesis}. First, our package supports
\texttt{library(recipes)}-style preprocessing steps. To use these, users
can specify preprocessing functions called \texttt{steps}, which can be
associated with different variables. For example, below we apply a
Yeo-Johnson transformation to \texttt{inctot}:

\begin{verbatim}
# specify a basic recipe in `library(recipes)`-style syntax
custom_recipe <- function(recipe) {
 
  recipe |>
    recipes::step_YeoJohnson(
      recipes::all_outcomes(), 
      id = "outcome yj", 
      skip = TRUE
    )
   
}

synth_spec_acs2 <- synth_spec_acs |> 
  update_custom_steps( # update our custom preprocessing steps...
      list(
        vars = c("inctot"), # for the `inctot` variable...
        steps = custom_recipe
      )  # use our custom recipe function
  )
\end{verbatim}

\textsubscript{Source:
\href{https://UI-Research.github.io/tidysynthesis-paper/tidysynthesis.qmd.html}{Article
Notebook}}

Next, our library supports additional randomization-based
post-processing of samples drawn from sequential models. Each process is
represented by a \texttt{noise()} object. Below, we define a function
that adds Gaussian noise with a fixed variance to each element of a
prediction vector:

\begin{verbatim}
# specify a `noise` object
reg_noise <- noise(
  add_noise = TRUE, 
  noise_func = add_noise_gaussian, # add Gaussian noise to the selected variables
  variance = 10  # ... with variance = 10
)

reg_noise
\end{verbatim}

\begin{verbatim}
Noise 
add_noise: TRUE
\end{verbatim}

\textsubscript{Source:
\href{https://UI-Research.github.io/tidysynthesis-paper/tidysynthesis.qmd.html}{Article
Notebook}}

Noise mechanisms can also depend on the confidential data and can be
defined in \texttt{noise\_func} with arbitrary user-defined functions.
We can then associate any noise mechanism with individual variables or
default behavior:

\begin{verbatim}
# add gaussian noise to all predicted numeric values
synth_spec_acs2 <- synth_spec_acs2 |> 
  update_synth_spec(
    # by default, use our noise mechanism for every regression variable
    default_regression_noise = reg_noise
  )
\end{verbatim}

\textsubscript{Source:
\href{https://UI-Research.github.io/tidysynthesis-paper/tidysynthesis.qmd.html}{Article
Notebook}}

We demonstrate the \texttt{constraints()} functionality in the snippet
below. For each numeric variable we synthesize, we can enforce minima
and maxima which may depend conditionally on other sequential
attributes. To do this, constraints are defined by creating a
\texttt{data.frame} which lists, one per row, variable names, numeric
minima and maxima, and a string representing a \texttt{tidyselect}-style
data-masking expression. Below is an example constraint specification.

\begin{verbatim}
# specify a collection of numeric constraints
numeric_constraints <- tibble::tribble(
  ~var, ~min, ~max, ~conditions,
  "age", 0, 95, "TRUE",  # all `age` values are between 0 and 95
  "inctot", 0, Inf, "TRUE", # all `inctot` values are non-negative
) 
\end{verbatim}

\textsubscript{Source:
\href{https://UI-Research.github.io/tidysynthesis-paper/tidysynthesis.qmd.html}{Article
Notebook}}

Similar behavior exists for constraining categorical values by
specifying either allowed or forbidden categorical levels.

To specify how constraints are enforced, users can specify
\texttt{max\_z\_num} and \texttt{max\_z\_cat}, which work analogously.
This parameter controls two different methods for enforcing constraints.
When \texttt{max\_z\_num\ \textgreater{}\ 0}, constraints are enforced
via rejection sampling. If a newly sampled record fails to satisfy the
constraints, new values are drawn until the constraints are satisfied;
this rejection sampling loop occurs up to \texttt{max\_z\_num} times.
After the \texttt{max\_z\_num}th attempt (or if
\texttt{max\_z\_num\ ==\ 0}), each synthetic value which fails to meet
the constraints is automatically post-processed to conform to the
constraint. For numeric variables, we set each failing record to the
closest of either the minimum or maximum value for each constraint (an
analogous process occurs with categorical variables). Below we show how
to update an existing \texttt{roadmap} instance to include constraints:

\begin{verbatim}
# add constraints to our `roadmap`
roadmap_acs2 <- roadmap_acs |>
  update_constraints(
    constraints_df_num = numeric_constraints, # add constraint data.frame
    max_z_num = 0  # enforce constraints without rejection sampling 
  )

roadmap_acs2[["constraints"]]
\end{verbatim}

\begin{verbatim}
Numeric constraints specified per variable: 
age: 1
famsize: 1
transit_time: 1
inctot: 1
Categorical constraints specified per variable: 
empstat: 0
\end{verbatim}

\textsubscript{Source:
\href{https://UI-Research.github.io/tidysynthesis-paper/tidysynthesis.qmd.html}{Article
Notebook}}

Finally, we use our new \texttt{roadmap} and \texttt{synth\_spec} in a
new synthesis.

\begin{verbatim}
# specify a new `presynth` with our newly modified `roadmap` and `synth_spec`
presynth2 <- presynth(
  roadmap = roadmap_acs2,
  synth_spec = synth_spec_acs2
)
\end{verbatim}

\textsubscript{Source:
\href{https://UI-Research.github.io/tidysynthesis-paper/tidysynthesis.qmd.html}{Article
Notebook}}

\begin{verbatim}
# synthesize the new `presynth`
set.seed(20251001)
result2 <- synthesize(presynth = presynth2, progress = TRUE)
\end{verbatim}

\textsubscript{Source:
\href{https://UI-Research.github.io/tidysynthesis-paper/tidysynthesis.qmd.html}{Article
Notebook}}

\begin{verbatim}
glimpse(result2[["synthetic_data"]])
\end{verbatim}

\begin{verbatim}
Rows: 500
Columns: 10
$ county       <fct> Lancaster, Sarpy, Other, Douglas, Other, Other, Other, Do~
$ gq           <fct> Household, Household, Household, Household, Household, Ho~
$ sex          <fct> Female, Female, Male, Male, Female, Female, Female, Male,~
$ marst        <fct> Married, Separated, Married, Single, Married, Single, Sep~
$ famsize      <dbl> 2.9870260, 6.2067210, 4.5708521, 1.8858135, -0.1791913, 7~
$ age          <dbl> 84.936584, 39.736101, 47.861174, 0.000000, 19.879675, 4.2~
$ inctot_NA    <fct> nonmissing value, nonmissing value, nonmissing value, mis~
$ inctot       <dbl> 9.954765e+03, 7.500340e+04, 6.300023e+04, NA, 6.004683e+0~
$ transit_time <dbl> 2.6596791, 21.9907233, 4.0682338, 0.5552695, 8.2226767, 1~
$ empstat      <fct> Employed, Employed, Employed, Employed, Employed, Employe~
\end{verbatim}

\textsubscript{Source:
\href{https://UI-Research.github.io/tidysynthesis-paper/tidysynthesis.qmd.html}{Article
Notebook}}

\subsection{Graceful Error Handling}\label{graceful-error-handling}

\texttt{tidysynthesis} catches many errors before running time-intensive
synthesis operations. For example, in the code below, we introduce a
data error and demonstrate how \texttt{presynth()} catches that error
before a user can call the time-intensive \texttt{synthesize()}
function.

\begin{verbatim}
# Introduce error: Assign variable famsize values that are 
# 1 (numeric) to "1" (character). This converts famsize  to a 
# categorical variable which conflicts with synth_spec_acs. 

acs_conf_data_with_error <- acs_conf_data |> 
  mutate(famsize = ifelse(famsize == 1, "1", famsize))

roadmap_with_error <- roadmap(
  conf_data = acs_conf_data_with_error,  # confidential data with error
  start_data = acs_start_data  # same start data as roadmap1
)
\end{verbatim}

\textsubscript{Source:
\href{https://UI-Research.github.io/tidysynthesis-paper/tidysynthesis.qmd.html}{Article
Notebook}}

\begin{verbatim}
# presynth catches error before synthesis
presynth_with_error <- presynth(
  roadmap = roadmap_with_error, 
  synth_spec = synth_spec_acs
)
\end{verbatim}

\begin{verbatim}
Error in .validate_presynth(presynth): Variable types in visit_sequence do not
match model types in synth_algorithms Problem variable(s): famsize
\end{verbatim}

\textsubscript{Source:
\href{https://UI-Research.github.io/tidysynthesis-paper/tidysynthesis.qmd.html}{Article
Notebook}}

\texttt{tidysynthesis} also provides useful functionality to extract
partially completed synthesis objects. This can be helpful for a user to
debug a more complicated code-related error or continue running a job
that failed for reasons unrelated to the code (e.g., a server crashed).

The code below introduces a more subtle error not caught by the
\texttt{presynth()} that causes the synthesis to fail. It assigns the
\texttt{age} variable to be sampled using \texttt{sample\_ranger}, a
custom sampler associated with random forest models using the
\texttt{ranger} package \citep{ranger}. Since \texttt{age} was created
using a decision tree from the \texttt{rpart} package
\citep{therneau2022}, the correct sampler for it would be
\texttt{sample\_rpart}.

\begin{verbatim}
# Create the erroneous synth_spec: 
synth_spec_with_error <- synth_spec(
  default_regression_model = rpart_mod_num, 
  default_classification_model = rpart_mod_cat, 
  default_regression_sampler = sample_rpart,
  default_classification_sampler = sample_rpart,
  
  # add a custom sampler associated with the ranger package
  # for the age variable. Age was synthesized using a 
  # decision tree from rpart, so it should have a sample_rpart sampler.
  custom_samplers = list(
    list(vars = c("age"), sampler = sample_ranger)
  )
)

presynth_with_error <- presynth(
  roadmap = roadmap_acs,
  synth_spec = synth_spec_with_error
)
\end{verbatim}

\begin{verbatim}
# synthesize fails but continues to run: 
result3 <- synthesize(
  presynth_with_error,
  
  # When set to TRUE (default is FALSE) keep roadmap, synth_spec, 
  # and workflow objects. This must be TRUE to recover the roadmap 
  # and synth_spec. 
  keep_workflows = TRUE,
                      
  # This also must be set to true for synthesize() to recover the 
  # roadmap and synth_spec. 
  keep_partial = TRUE
)
\end{verbatim}

\textsubscript{Source:
\href{https://UI-Research.github.io/tidysynthesis-paper/tidysynthesis.qmd.html}{Article
Notebook}}

Now, we call \texttt{postsynth\_to\_roadmap()} and
\texttt{postsynth\_to\_synth\_spec()} to recover the partially completed
\texttt{roadmap} and \texttt{synth\_spec} objects. Both are updated so
that the remaining variables to synthesize and the specifications to
synthesize them only apply to variables that have not yet been
synthesized. As an example, note how the new visit sequence does not
include \texttt{famsize}, the first variable in the visit sequence which
was synthesized correctly, and instead begins with \texttt{age}, the
variable that caused an error in the synthesis. The objects returned
from these functions can be used in subsequent synthesis operations and
to explore issues that caused the failure.

\begin{verbatim}
recovered_roadmap <- postsynth_to_roadmap(result3)
recovered_synth_spec <- postsynth_to_synth_spec(result3)

recovered_roadmap
\end{verbatim}

\begin{verbatim}
Roadmap: 
conf_data: 1500 observations, 10 variables 
start_data: 500 observations, 5 variables
\end{verbatim}

\begin{verbatim}
recovered_roadmap$visit_sequence$visit_sequence
\end{verbatim}

\begin{verbatim}
[1] "age"          "inctot_NA"    "inctot"       "transit_time" "empstat"     
\end{verbatim}

\begin{verbatim}
recovered_synth_spec
\end{verbatim}

\begin{verbatim}
Synthesis specification with user-specified components: 
* default_regression_model
* default_classification_model
* default_regression_sampler
* default_classification_sampler
* custom_samplers
\end{verbatim}

\textsubscript{Source:
\href{https://UI-Research.github.io/tidysynthesis-paper/tidysynthesis.qmd.html}{Article
Notebook}}

\section{Conclusion}\label{sec-conclusion}

In this paper, we introduced \texttt{tidysynthesis}, an R package which
enables users to perform generalized sequential synthesis on tabular
data, filling a gap in the existing synthetic software generation
landscape. It utilizes a modular approach to allow users to easily
configure a variety of synthetic data specifications. As a meta-package,
\texttt{tidysynthesis} users can harness the full range of features in
the \texttt{tidymodels} and \texttt{tidyverse} data science library
ecosystem, including feature- and target-engineering functions and
predictive models from \texttt{tidymodels}. It also offers numerous
synthetic-data-specific capabilities, like the ability to integrate
mid-synthesis constraints and infuse estimates with randomized noise. We
also demonstrated the power of \texttt{tidysynthesis} by synthesizing a
non-trivial set of data from the 2019 American Community Survey.

\texttt{tidysynthesis} is under continuous development and currently
supports a wide variety of use cases, ranging from education and
technical training to large-scale synthetic data generation projects.
Interested readers can follow and contribute to the project on
GitHub\footnote{https://github.com/UrbanInstitute/tidysynthesis} and
consult the \texttt{tidysynthesis} documentation website\footnote{https://ui-research.github.io/tidysynthesis-documentation/}.
As the data privacy and synthetic data research communities continue to
produce synthesis generation methods, we hope these communities will
both use and contribute to the package to enable effective
implementation decision-making across methods. Overall, we see
\texttt{tidysynthesis} as providing much-needed software infrastructure
to bridge the gap between synthetic data methods research and the needs
of practitioners.

\newpage

\section*{Author Contributions}\label{author-contributions}
\addcontentsline{toc}{section}{Author Contributions}

We used the CRediT taxonomy to indicate author contributions to this
paper. See website to learn more about the CRediT taxonomy,
https://credit.niso.org.

\begin{itemize}
\tightlist
\item
  \textbf{ARW:} Conceptualization, Funding acquisition, Investigation,
  Methodology, Project Administration, Software, Supervision,
  Validation, Visualization, Writing -- original draft, Writing --
  review \& editing
\item
  \textbf{JS:} Data Curation, Investigation, Methodology, Project
  Administration, Resources, Software, Supervision, Validation,
  Visualization, Writing -- original draft, Writing -- review \& editing
\item
  \textbf{GM:} Investigation, Methodology, Software, Validation,
  Visualization, Writing - review and editing, Writing - original draft
\end{itemize}

\section*{Acknowledgement}\label{acknowledgement}
\addcontentsline{toc}{section}{Acknowledgement}

Kyle Ueyama, Maddie Pickens, Livia Mucciolo, Noah Zwiefel, Philip
Stallworth, Ajjit Narayanan, Nikhita Airi, Elyse McFalls, and Claire
Morton all made essential contributions to \texttt{tidysynthesis}.
Thanks to Claire McKay Bowen for securing financial support for the
\texttt{tidysynthesis} package development and this publication.

\texttt{tidysynthesis} was funded by the Alfred P. Sloan Foundation
{[}G-2022-17149 and G-2020-14024{]} and National Science Foundation
National Center for Science and Engineering Statistics
{[}49100422C0008{]}.

\renewcommand\refname{References}
\bibliography{references.bib}

\end{document}